\documentclass[conference]{IEEEtran}
\IEEEoverridecommandlockouts
\usepackage{cite}
\usepackage{amsmath,amssymb,amsfonts}
\usepackage{algorithmic}
\usepackage{graphicx}
\usepackage{textcomp}
\usepackage{xcolor}
\usepackage{multirow}
\usepackage{fancyhdr}
\usepackage{float}
\usepackage{subfig}
\def\BibTeX{{\rm B\kern-.05em{\sc i\kern-.025em b}\kern-.08em
    T\kern-.1667em\lower.7ex\hbox{E}\kern-.125emX}}

\renewcommand{\baselinestretch}{0.95}
\begin{document}

\title{\textbf{QForce-RL}: \textbf{Q}uantized \textbf{F}PGA-\textbf{O}ptimized \textbf{R}einforcement Learning \textbf{C}ompute \textbf{E}ngine\\}

\author{
Anushka Jha\textsuperscript{$\ast$}, Tanushree Dewangan\textsuperscript{$\ast$}, 
Mukul Lokhande,
Santosh Kumar Vishvakarma,~\IEEEmembership{Senior Member, IEEE,} \\
Both authors\textsuperscript{$\ast$} have contributed equally to this work.\\

        % <-this % stops a space
        
\thanks{
This work was supported by the Special Manpower Development Program for Chip to Startup (SMDP-C2S), Ministry of Electronics and Information Technology (MeitY), Govt. Of India. 
The authors are associated with the NSDCS Research Group, Department of Electrical Engineering, IIT Indore, Simrol-453552, India.
\textbf{Corresponding author}: Santosh Kumar Vishvakarma, \textbf{E-mail:} skvishvakarma@iiti.ac.in.}% <-this % stops a space
% \thanks{Manuscript received ; revised .}
}

% \thanks{This work was supported by the Special Manpower Development Program for Chip to Startup (SMDP-C2S), Ministry of Electronics and Information Technology (MeitY), Govt. Of India.\\
% Both the    authors marked \textsuperscript{$\ast$}, and \textsuperscript{$\alpha$} have contributed equally to this work.
% }
% }

% \author{\IEEEauthorblockN{Anushka Jha\textsuperscript{$\ast$}}
% \IEEEauthorblockA{\textit{Dept. of Electrical Engineering} \\
% \textit{Indian Institute of Technology }\\
% Indore, India \\
% ee220002013@iiti.ac.in}
% \and
% \IEEEauthorblockN{Tanushree Dewangan\textsuperscript{$\ast$}}
% \IEEEauthorblockA{\textit{Dept. of Electrical Engineering} \\
% \textit{Indian Institute of Technology }\\
% Indore, India \\
% ee220002077@iiti.ac.in}
% \and

% \IEEEauthorblockN{Mukul Lokhande}
% \IEEEauthorblockA{\textit{Dept. of Electrical Engineering} \\
% \textit{Indian Institute of Technology }\\
% Indore, India \\
% phd2201102020@iiti.ac.in}
% \and

% \IEEEauthorblockN{Akash Sankhe\textsuperscript{$\alpha$}}
% \IEEEauthorblockA{\textit{Dept. of Electrical Engineering} \\
% \textit{Indian Institute of Technology }\\
% Indore, India \\
% ms2304102003@iiti.ac.in}
% \and

% \IEEEauthorblockN{Omkar Kokane\textsuperscript{$\alpha$}}
% \IEEEauthorblockA{\textit{Dept. of Electrical Engineering} \\
% \textit{Indian Institute of Technology }\\
% Indore, India \\
% mt2302102023@iiti.ac.in}
% \and

% \IEEEauthorblockN{Santosh Kumar Vishvakarma}
% \IEEEauthorblockA{\textit{Dept. of Electrical Engineering} \\
% \textit{Indian Institute of Technology }\\
% Indore, India \\
% skvishvakarma@iiti.ac.in}
% }

\maketitle
\thispagestyle{fancy}

\begin{abstract}
Reinforcement Learning (RL) has outperformed other counterparts in sequential decision-making and dynamic environment control. However, FPGA deployment is significantly resource-expensive, as associated with large number of computations in training agents with high-quality images and possess new challenges. In this work, we propose QForce-RL takes benefits of quantization to enhance throughput and reduce energy footprint with light-weight RL architecture, without significant performance degradation. QForce-RL takes advantages from E2HRL to reduce overall RL actions to learn desired policy and QuaRL for quantization based SIMD for hardware acceleration. We have also provided detailed analysis for different RL environments, with emphasis on model
size, parameters, and accelerated compute ops. The architecture is scalable for resource-constrained devices and provide parametrized efficient deployment with flexibility in latency, throughput, power, and energy efficiency. The proposed QForce-RL provides performance enhancement up to 2.3 $\times$ and better FPS - 2.6 $\times$ compared to SoTA works. 

\end{abstract}


\section{Introduction}

\IEEEPARstart{A}{rtificial} Intelligence (AI) has became an integral part of human life, with quick assistance for key data-driven decision-making. However the hardware design efforts efforts, cost have risen exponentially with the decreasing node size. Generative AI assisted by deep Reinforcement Learning (RL)\cite{QuaRL, E2HRL}, plays significant role to reduce the average design cost/chip size, days contributed by every person/chip size. For example, 60\% of chip-designer's efforts can be saved with automation and effectively reducing overall design cycle time, which otherwise was wasted in debug or checklist-related tasks across diverse tasks such as tool usage, design specs, test-bench creation, and root cause analysis of flows\cite{NVIDIA-chipnemo}. The more details can be understood from Table \ref{comp-ai-benefits}.

RL play pivotal role in design space exploration, enabling iterative optimization of power, performance, and area (PPA) through trial-and-error learning. RL models adapt dynamically to evolving constraints, effectively guiding designers for optimal choices for layout, synthesis, and verification stage. For instance, the floor-planing tool by Google DeepMind\cite{google-alphachip} has outperformed human experts and shortened design cycles and improving first-pass silicon success rates. However, the increasing use of AI has significantly raised compute demands, from DNNs, LLMs to RL. This underscores the need for scalable and configurable AI accelerators to effectively serve/train models within constrained memory and compute budget, such as Google AlphaGo, Apple Vision Pro and NVIDIA GROOT\cite{alphago}. Thus, achieving high compute throughput is need of the hour while addressing critical bottlenecks such as memory usage, data bandwidth, and access latency, both within and across chips.

\begin{table}[]
\caption{AI Design Improvements via RL}
\centering
\vspace{-2mm}
\label{comp-ai-benefits}
\resizebox{0.9\columnwidth}{!}{%
\begin{tabular}{c|c|c|c}
\hline
\textbf{Design} & \textbf{\begin{tabular}[c]{@{}c@{}} Improved (X)\\ Productivity\end{tabular}} & \textbf{Parameter} & \textbf{Benefits} \\ \hline
AI Processor 5nm & 6 & Total Power & 7.7\% \\ \hline
GPU Server 5nm & 8 & PPA & 5\%+ \\ \hline
Mobile SoC (1B gates) & 60 & Debug Effort & Reduction \\ \hline
Automotive SoC & 6 & Change analysis & Faster \\ \hline
Mobile CPU 3nm & 5 & Total Power & 12.5\% \\ \hline
Server SoC 4nm & 5 & Leakage Power & 28\% \\ \hline
Automotive 7nm & 5 & Leakage Power & 22\% \\ \hline
TPU PCB & 50 & Wire Length & 14\% \\ \hline
Auto PCB \& Package & 30 & Return loss & 134\% \\ \hline
\end{tabular}}
\vspace{-5mm}
\end{table}

TOPS performance has improved nearly 60,000 $\times$ in past two decades, while DRAM and interconnect bandwidths have only scaled by 30$\times$ and 100$\times$, respectively. This disparity emphasizes on critical need for enhanced parallel compute architectures - especially for RL-based AI hardware, which has not addressed in state-of-the-art (SoTA) approaches~\cite{Survey-RL, RL-Mobileaccl, RL-survey, RL-nature}. Prior works explored pre-training and finetuning to address the increasing complexity of software frameworks, while recent efforts, DeepSeek\cite{deepseek} shows compact models combined with quantized-RL and distillation boost performance. Prior work, E2HRL~\cite{E2HRL} have predominantly used 32-bit precision with hierarchical designs, however lack support for SIMD compute commonly adopted in DNN hardware\cite{flex-PE}. Furthermore, prior work hasn't addressed diverse AF requirements in RL workloads (e.g., sigmoid/tanh in LSTM, ReLU in FC layers)~\cite{E2HRL}, which could be derived from application specific derivations of CORDIC or Taylor-based methods\cite{flex-PE, Flex-SFU}. Our proposed approach leverages quantization and reconfigurability to deliver a lightweight yet powerful RL compute engine. This work marks a significant step toward hardware-software co-design for efficient and scalable RL implementation.

\begin{figure}
    \centering
    \includegraphics[width=0.8\linewidth]{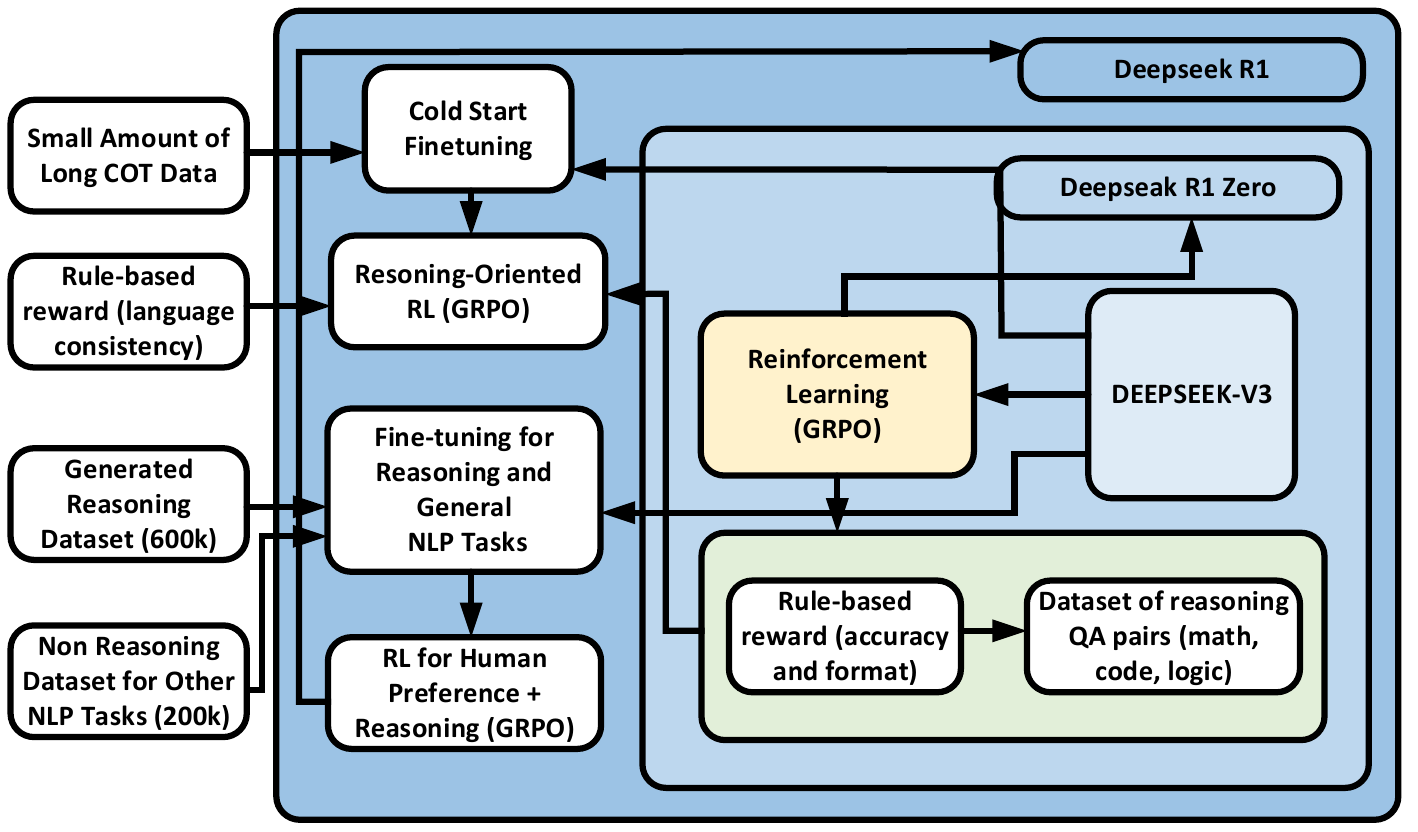}
    \vspace{-2mm}
    \caption{Deepseek model, showing emphasis on RL}
    \label{deepseek}
    \vspace{-5mm}
\end{figure}
 
% The work proposes Q-MAC, which supports runtime flexibility between multi-precision (8/16/32) operations at improved throughput. The work incorporates versatile area-efficient V-ACT for support across diverse NAFs such as Tanh, ReLU, SoftMax, and Sigmoid with SIMD benefits. The runtime precision switching contributes significantly to resource savings while minimizing memory footprint. Combinedly, these advancements enable effective RL accelerators for edge-AI devices for quick event-triggered response, superior energy efficiency, and flexible programmable solutions. We have analysed the software-based implementation platform and adaptive-modular RL accelerator architecture on FPGA and provided details of implementation.  

The major contributions of this work are:

\begin{itemize}

    \item \textbf{SIMD Multi-precision Q-MAC:} We propose a configurable Q-MAC unit supporting FxP8/16/32 precisions with SIMD parallelism (16/4/1 MACs per cycle). The design emphasizes resource reuse, reducing memory footprint while enhancing throughput for RL workloads via parallelized multiplication blocks.

    \item \textbf{Versatile CORDIC-based activation function (V-ACT):} A reconfigurable, low-latency V-ACT supporting Tanh, Sigmoid, ReLU, and Softmax at SIMD precision (FxP8/16/32), personalized for FC and LSTM layers in RL.

    \item \textbf{Analysis for Hierarchical Deep RL Accelerator:} We present a scalable and adaptive RL accelerator evaluated on FPGA with quantized HRL policy. Our empirical analysis shows significant improvements up to 2.6× throughput and reduced resource utilization, making it suitable for resource-constrained execution.
    
\end{itemize}

The structure of this paper is as follows:
Section \ref{sec2} discusses the quantization strategy for RL and its impact on performance.
Section \ref{sec3} presents the QForce-RL architecture, highlighting its reconfigurability and precision-adaptive SIMD design. It also details the implementation methodology and provides a comprehensive performance evaluation.
Finally, Section \ref{sec4} concludes the paper and outlines future directions.

\section{Need for QForce-RL and Possible trade-offs}
\label{sec2}

RL involves extensive arithmetic operations, particularly in sequential decision-making and reward accumulation, demanding significantly higher computational resources than traditional deep learning. While GPUs can handle multi-precision operations through type-casting, Custom compute engines (CEs) are more desirable for delivering consistent performance across diverse application scenarios. At the core of many AI workloads lies matrix algebra, which underpins applications such as autonomous navigation (robots, drones, vehicles), virtual/augmented reality, and machine learning. Multiply-accumulate (MAC) units are critical for efficiently performing matrix operations. Notably, as models have scaled from VGG-16 (140M parameters) to GPT-2 (175B parameters), MAC operation demands have surged from 15.5 billion to 42 billion, emphasizing on the need for hardware accelerators to support scalable and compute-intensive AI workloads\cite{flex-PE, lpre}.

Quantization plays pivotal role for high parallel throughput at low precision, enabling enhanced hardware performance without compromising application accuracy. Modern industry-grade ASICs increasingly support a wide spectrum of numerical formats to balance performance and precision trade-offs. For example, NVIDIA's GB200 (Tensor Core v4) supports data types ranging from FP64 to INT2, including FP32/16/8/6/4, TF32, BF16. Similarly, Qualcomm's AI100 supports FP64/32/16, TF32, BF16, and INT8, while Intel Xeon processors offer extensive format compatibility, including FP32/16/8, INT32/16, TF32, BF16, and ultra-low-bit formats like U8/S8, U4/S4, and U2/S2. Similar to commercial trends, a reconfigurable SIMD processing element (PE) must efficiently support a wide range of precision levels for RL workloads as well, to enable high throughput at low energy per operation for resource-constrained AI deployments.

\begin{figure}
    \centering
    \includegraphics[width=0.8\linewidth]{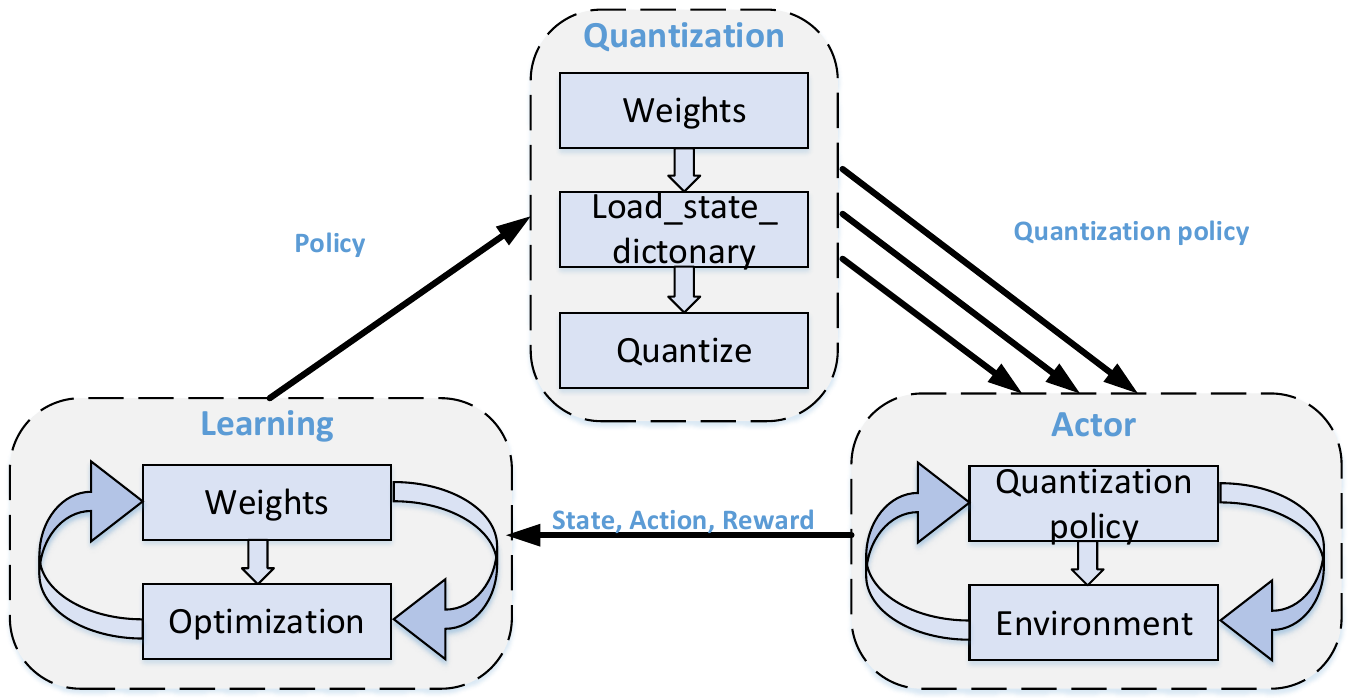}
    \caption{Q-Actor Software emulation setup (RL)}
    \label{RL-Quant}
    \vspace{-5mm}
\end{figure}

Q-Actor\cite{QuaRL} is a novel software-emulated reinforcement learning (RL) framework designed to accelerate distributed actor-learner training. It leveraged INT 8-bit quantized actors, significantly speeding up data collection, achieving end-to-end speedups ranging from 1.4 $\times$ to 5.6 $\times$, without impacting learning convergence. Building upon this, adaptive fixed-point 8-bit (AdFxP8) formats have emerged, offering improved accuracy and energy efficiency over conventional INT8 while utilizing the same hardware. This enables better hardware utilization and reduces communication overhead in distributed RL training. Workload characterization studies indicate that actor policy inference dominates computational time, followed by the learner’s gradient computation, model updates, and actor-learner synchronization. This motivates the need for quantized actor inference optimization. Unlike DNNs, RL introduces a feedback loop with the environment, which makes its quantization behavior unique. Although quantization errors may propagate through state transitions, RL policies are inherently resilient with feedback enabling recovery from quantization-induced errors. It is also important to note that conventional QAT provides little training speedup unless the hardware and runtime libraries natively support low-precision operations. Our approach builds upon these insights, aligning with prior work in Q-Actor and E2HRL~\cite{QuaRL, E2HRL}.

\begin{figure}[!b]
\vspace{-5 mm}
    \centering
    \subfloat[]{\includegraphics[width=0.5\columnwidth]{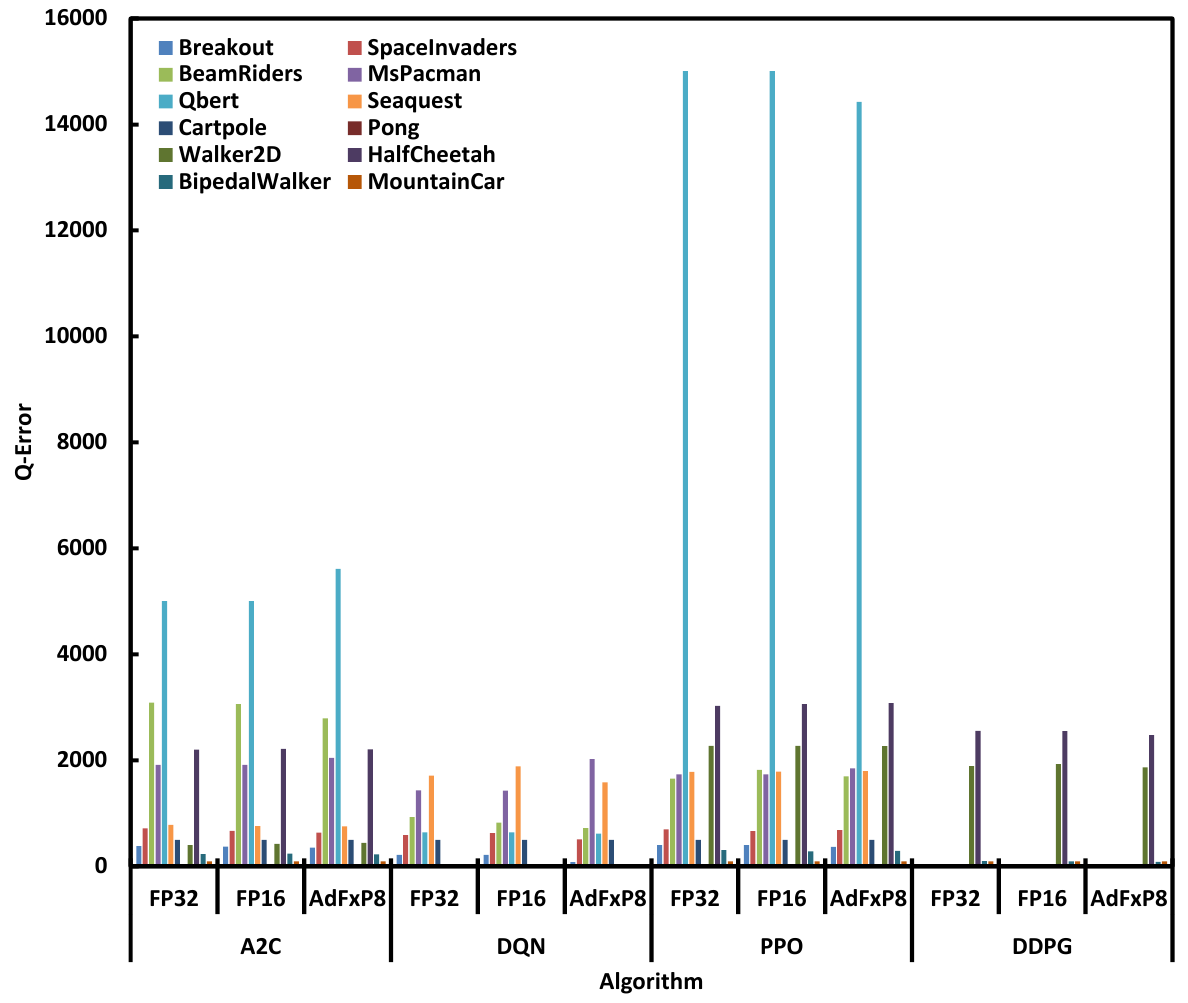}
    \label{rl_perf}}
    %\hfill
    \centering
    \subfloat[]{\includegraphics[width=0.40\columnwidth, height=35mm]{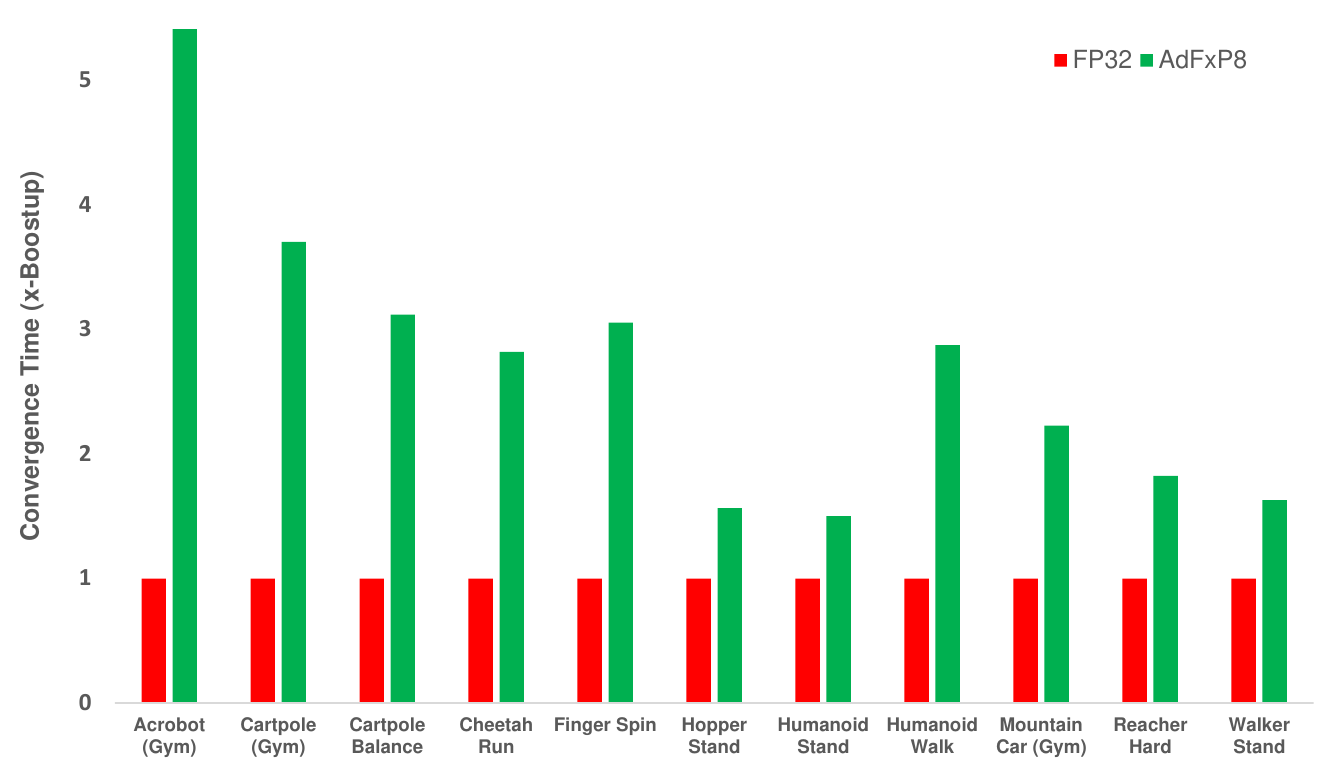}
    \label{Qactor_toc}}
    \hspace{0.1\columnwidth}
    \caption{(a) Application performance comparison (rewards) for A2C, DQN, PPO, and DDPG algorithm, comparison with FP32-baseline\cite{QuaRL}, (b) Off-chip Components.}
\end{figure}

The Q-Actor policy leverages quantization to accelerate experience collection and reduce communication overhead between the actor and learner, thereby facilitating faster RL convergence. The Q8 policy exhibits minimal to no degradation in reward performance, enabling significant end-to-end speedups across various RL algorithms as shown in Fig \ref{Qactor_toc}. There is scope for exploring mixed-precision quantization, where different components of the RL workload such as the policy network, value estimator, or embedding layers operate at varying bit-widths. This approach can further optimize memory usage, minimize data transfer, and enhance the utilization of SIMD compute resources. In reinforcement learning, training relies on experience generation through repeated policy execution in a dynamic environment, along with continuous actor-learner communication for policy updates and performance refinement. Thus, precision-aware design becomes crucial for improving both computational efficiency and learning convergence.

\begin{figure}[!t]
    \centering
    \includegraphics[width=0.85\columnwidth, height=22.5mm]{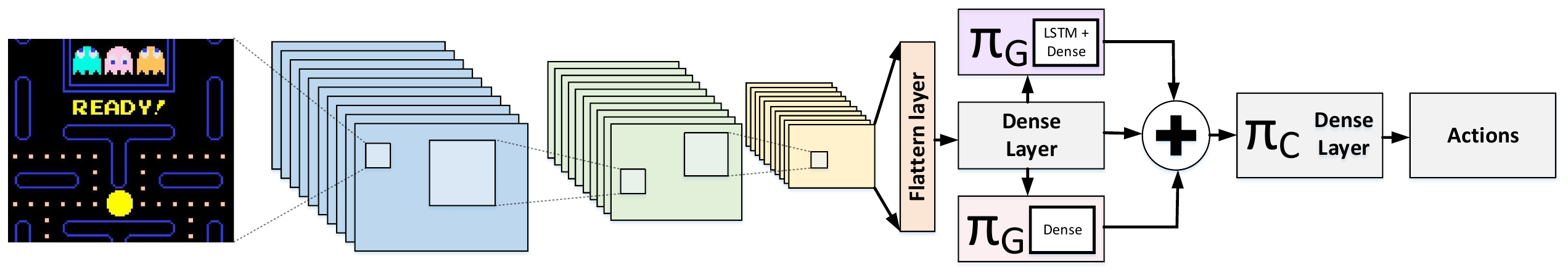}
    \caption{Q-HRL network architecture}
    \vspace{-5mm}
    \label{fig:qhrl_nw}
\end{figure}

The Q-Actor RL framework enables each actor to collect new training samples based on its current policy, which are then relayed to the learner. This facilitates continuous experience accumulation and policy refinement. To maintain consistency, periodic synchronization is performed between the actor and learner policies. Quantization significantly accelerating training and reducing communication overhead here. As illustrated in Fig.~\ref{RL-Quant}, the three core components of the system are the Actor, Learner, and Quantizer. Each actor performs rollouts using a randomly initialized policy, broadcasting the environment state, action, and reward to the learner, which in turn updates the policy based on this feedback. The inference latency is largely coming from neural network used within the actor for action generation. The uniform affine quantization approach\cite{QuaRL} can be modeled as:
\begin{equation}
     Q_n(W_q) = \operatorname{round} \left( \frac{W_{fp32} \times {2^n}} {(| \min(W_q,0) | + | \max(W_q,0) |)} \right)
\end{equation} 

Experimental results indicate that the Q-RL policy incur a negligible reward loss compared to the FP32-based RL policy. Additionally the architecture yields \textbf{O(n)} hardware savings across \( n \) actors. The Q-FxP8 computations further enhance execution speed and energy efficiency, making them ideal for distributed RL scenarios.

\section{QForce-RL architecture \& Analysis}
\label{sec3}

\begin{figure}
    \centering
    \includegraphics[width=0.85\linewidth]{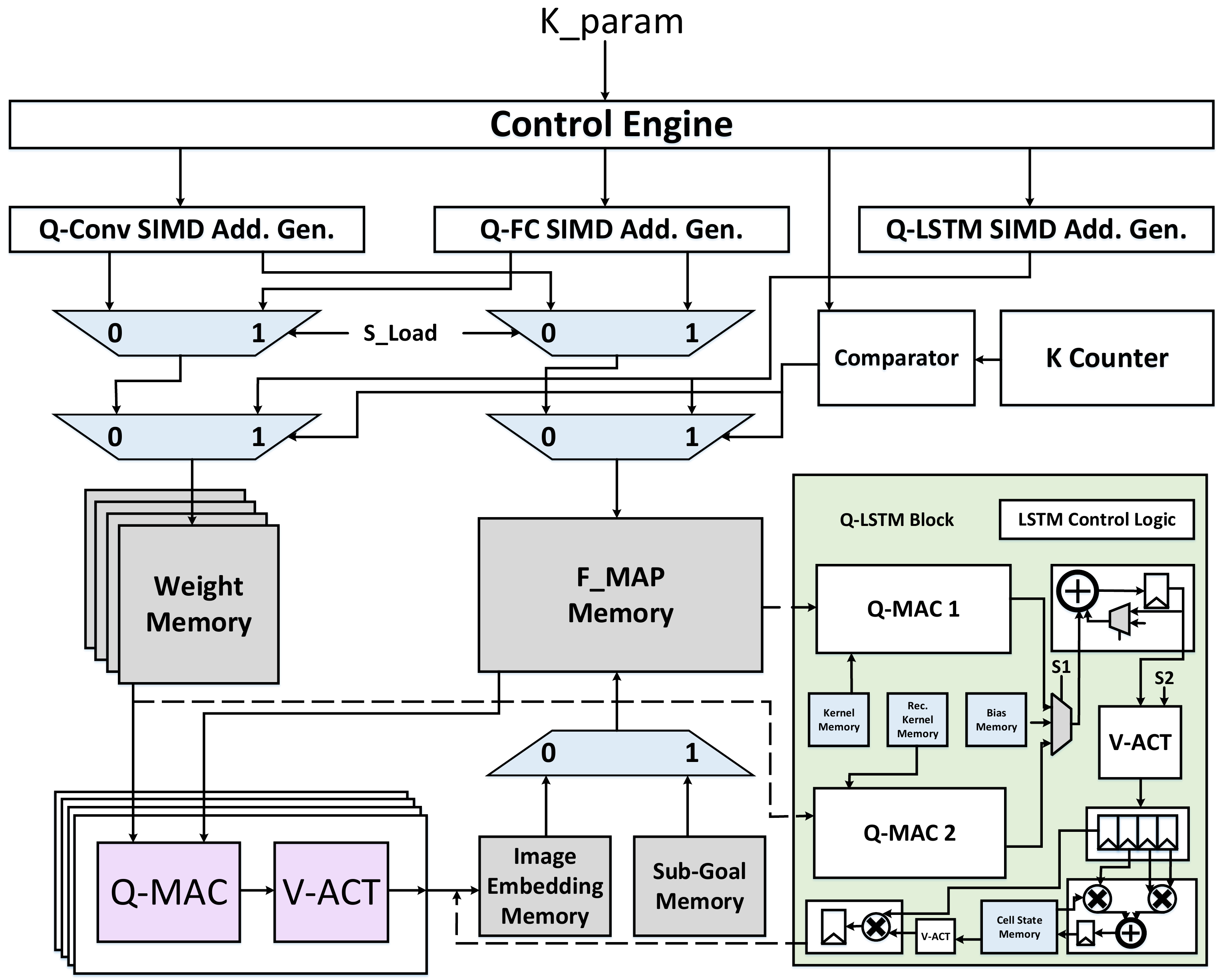}
    \caption{QForce-RL architecture}
    \label{fig:QForce-RL}
    \vspace{-7mm}
\end{figure}

The proposed QForce-RL architecture as shown in Fig.~\ref{fig:QForce-RL} comprises quantized Q-Conv, Q-LSTM, fully connected (Q-FC), and Softmax layers. The agent includes two modules:sub-goal and action - both of which operate on image embeddings extracted by Q-Conv layers. The sub-goal module generates an intermediate goal from the current observation, which, along with the original embedding, is processed by the action module to determine the final action. The design follows a two-stage training strategy based on proximal policy optimization (PPO), adapted from the conventional FxP32-based E2HRL framework. Once the action module is trained, its weights are frozen, and the sub-goal module is fine-tuned independently. The sub-goal path employs an MLP-style Q-FC block or a Q-LSTM layer to model temporal dependencies. The Q-LSTM, a specialized RNN variant~\cite{flex-PE}, retains long- and short-term context critical for time-dependent decision-making. The set of architectural choice to construct efficient models similar to E2HRL search space\cite{E2HRL}, enabling a balance between policy performance and hardware efficiency, while leveraging quantization to minimize memory and compute overhead.

The baseline E2HRL~\cite{E2HRL} comprises three Q-Conv layers, followed by a flattened output processed by a Q-FC layer to generate a 32-dimensional image embedding. This embedding is fed into the sub-goal module implemented using either a Q-FC or Q-LSTM block to compute the sub-goal vector. The resulting vector is concatenated with the original embedding and passed through a final Softmax layer to generate the agent's action. It was also identified that Q-FC\_2 and Q-LSTM\_4 to be energy-efficient and hardware-friendly at FxP32 precision. Our quantized implementation further reduces memory usage and computation, making it highly suitable for FPGA-based edge deployments with limited Block RAM. The architecture employs PE arrays capable of executing 4/8 Q-MACs per cycle, yielding throughput of 16/4/1 for FxP8/16/32, respectively. Dedicated memory blocks are allocated for weights, features, and intermediate activations, each with address generators tailored to the Q-Conv, Q-FC, and Q-LSTM layers.
An independent Q-LSTM block, controlled with hyperparameter  K, performs matrix-vector multiplications with Sigmoid/Tanh activations (via V-ACT). The memory hierarchy serves to store weights, intermediate features, image embeddings, and sub-goal vectors, and is structured to support parallelism across increasing PE counts, optimizing for scalability and resource efficiency.

At a given timestep $t$, the Long Short-Term Memory (LSTM) unit performs a series of gated operations to compute its internal state and output. First, the input gate $i_t$, forget gate $f_t$, and output gate $o_t$ are computed using the sigmoid activation function $\sigma(\cdot)$ applied to an affine transformation of the current input $x_t$ and the previous hidden state $h_{t-1}$. Specifically, $i_t = \sigma(W_{xi}x_t + W_{hi}h_{t-1} + b_i)$, $f_t = \sigma(W_{xf}x_t + W_{hf}h_{t-1} + b_f)$, and $o_t = \sigma(W_{xo}x_t + W_{ho}h_{t-1} + b_o)$. The candidate cell input $g_t$ is computed using the hyperbolic tangent function, $g_t = \tanh(W_{xg}x_t + W_{hg}h_{t-1} + b_g)$. The cell state $c_t$ is then updated by combining the previous cell state $c_{t-1}$ and the candidate input, modulated by the forget and input gates respectively: $c_t = f_t \odot c_{t-1} + i_t \odot g_t$, where $\odot$ denotes element-wise multiplication. Finally, the hidden state output $h_t$ is obtained by applying the $\tanh$ function to the updated cell state and modulating it with the output gate: $h_t = \tanh(c_t) \odot o_t$.

The hardware architecture comprises modular blocks, including a Q-Conv unit for 2D convolution with embedded MAC. Max-pooling is replaced with a stride of two, followed by a ReLU activation. The configurable Q-FC/Q-LSTM block executes either dense (FC-HRL) or LSTM (LSTM-HRL) operations and integrates two Q-MAC units, V-ACT modules, arithmetic logic, and dedicated memories for weights, recurrent kernels, biases, and cell states. A shared on-chip memory is used to store weights, feature maps, image embeddings, and sub-goal vectors. The hardware supports reconfigurability with key hyperparameters: \# PEs, filter sizes, Q-Conv \# filters, and Q-FC \# neurons. This enables runtime performance–area trade-offs, allowing flexible deployment across a range of RL applications and resource constraints.

\begin{figure}
    \centering
    \includegraphics[width=0.85\linewidth]{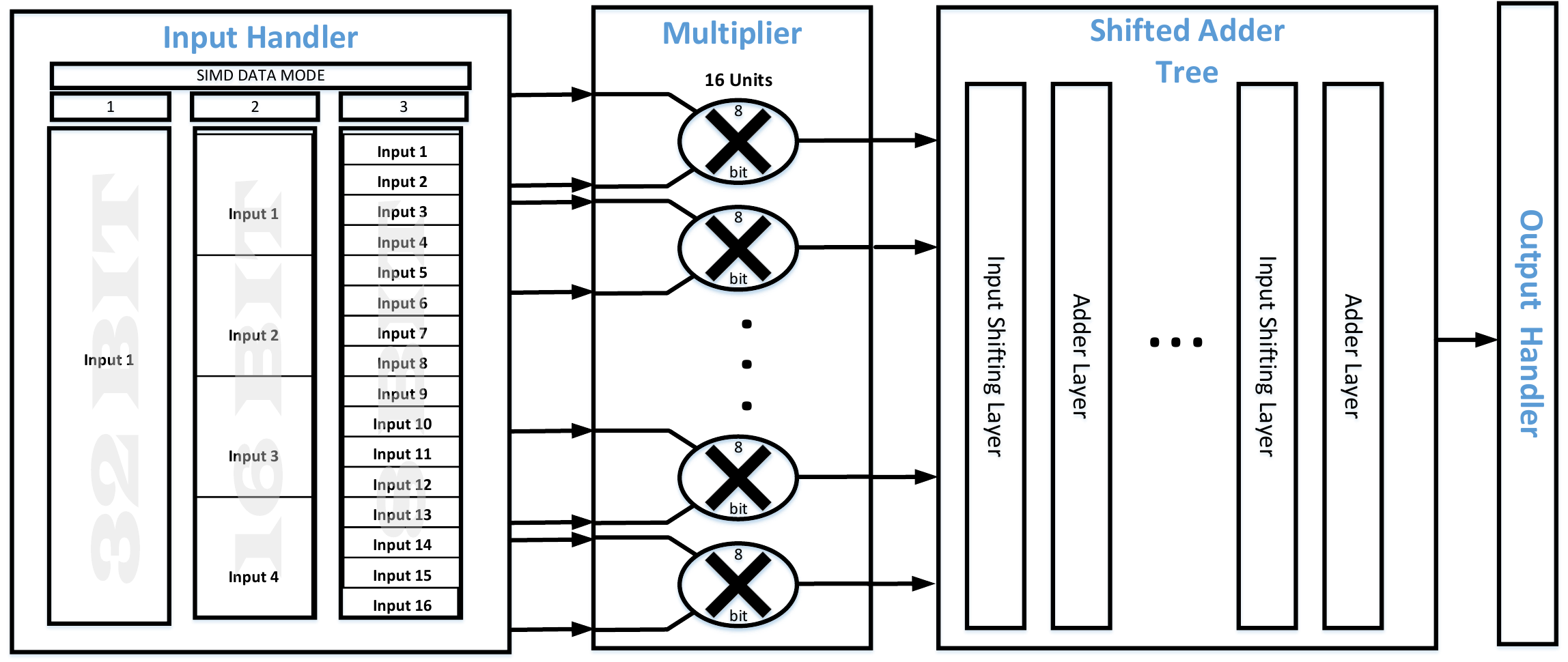}
    \caption{The proposed Q-MAC 8/16/32 design}
    \label{fig:Q-MAC}
    \vspace{-5mm}
\end{figure}

\begin{table}[!b]
\vspace{-5mm}
\caption{FPGA Resource  Utilization of SoTA and Q-MAC}
\label{fpga-mac}
\renewcommand{\arraystretch}{1.15}
\resizebox{\columnwidth}{!}{%
\begin{tabular}{|c|c|c|c|c|c|c|}
\hline
Design & FPGA & Precision & \begin{tabular}[c]{@{}c@{}}Freq. \\ (MHz)\end{tabular} & \begin{tabular}[c]{@{}c@{}}Power \\      (mW)\end{tabular} & LUT & FF \\ \hline
FP-RPE TCAS-II'24\cite{Li-TCASII'24} & VCU129 & FP-8/16/32 & 216.5 & 0.29 & 8054 & 1072 \\ \hline
FP-MPE\cite{FP-MPE} & VCU129 & FP-8/16/32 & 180 & 0.38 & 8065 & 1718 \\ \hline
Acc\_App\_MAC\cite{Acc-App-MAC-TCAD'22} & VC707 & 32-b & 140 & 8.5 & 1013 & NR \\ \hline
Quant-MAC\cite{QuantMAC} & VC707 & 16-b & 283 & 11.77 & 106 & 168 \\ \hline
Quant-MAC\cite{QuantMAC} & VC707 & 8-b & 350 & 6.36 & 52 & 88 \\ \hline
Unified-VMAC\cite{Unified-VMAC-TCASII'22} & VC707 & Posit/FP-8/16/32 & 267 & 8.34 & 5972 & 1634 \\ \hline
Quant-VMAC\cite{QuantMAC} & VC707 & FxP-8/16/32 & 250 & 21.34 & 1503 & 2417 \\ \hline
V-MAC & VC707 & FxP-8/16/32 & 190 & 35.2 & 2879 & 1076 \\ \hline
Proposed Q-MAC & VC707 & AdFxP-8/16/32 & 232 & 4.2 & 835 & 1062 \\ \hline
\end{tabular}}
\end{table}

\begin{table*}[!t]
\vspace{-5mm}
\caption{Comparative analysis of ASIC performance with SoTA MAC works using CMOS 28nm}
\centering
\renewcommand{\arraystretch}{1.15}
\resizebox{0.8\textwidth}{!}{%
\label{MAC-ASIC}
\begin{tabular}{|c|c|c|c|c|ccc|ccc|}
\hline
\multirow{2}{*}{\textbf{Pub.-Year}} & \multirow{2}{*}{\textbf{Precision}} & \multirow{2}{*}{\begin{tabular}[c]{@{}c@{}}\textbf{Power}\\(mW)\end{tabular}} & \multirow{2}{*}{\begin{tabular}[c]{@{}c@{}}\textbf{Area}\\(mm\textsuperscript{2})\end{tabular}} & \multirow{2}{*}{\begin{tabular}[c]{@{}c@{}}\textbf{Freq.}\\(GHz)\end{tabular}} & \multicolumn{3}{c|}{\textbf{Energy efficiency (GOPS/W)}} & \multicolumn{3}{c|}{\textbf{Compute Density (GOPS/mm\textsuperscript{2})}} \\ \cline{6-11} 
 &  &  &  &  & \multicolumn{1}{c|}{32-b} & \multicolumn{1}{c|}{16-b} & 8-b & \multicolumn{1}{c|}{32-b} & \multicolumn{1}{c|}{16-b} & 8-b \\ \hline
TCAS-II'24\cite{Tan-TCASII'24} & FP-64/32/16 & 72.3 & 0.022 & 1.56 & \multicolumn{1}{c|}{43.15} & \multicolumn{1}{c|}{86.31} & N/A & \multicolumn{1}{c|}{141.82} & \multicolumn{1}{c|}{283.64} & N/A \\ \hline
TCAD'24\cite{Tan-TCAD'24} & FP-64/32/16 & 82.4 & 0.024 & 1.47 & \multicolumn{1}{c|}{35.68} & \multicolumn{1}{c|}{71.36} & N/A & \multicolumn{1}{c|}{122.50} & \multicolumn{1}{c|}{245} & N/A \\ \hline
TCAS-II'22\cite{Unified-VMAC-TCASII'22} & Posit/FP-32/16/8 & 99 & 0.050 & 0.67 & \multicolumn{1}{c|}{13.54} & \multicolumn{1}{c|}{27.07} & 54.14 & \multicolumn{1}{c|}{26.80} & \multicolumn{1}{c|}{53.60} & 107.20 \\ \hline
TCAS-II'22\cite{Li-TCASII'22} & FP-32/8/BF-16 & 43.8 & 0.071 & 0.67 & \multicolumn{1}{c|}{30.59} & \multicolumn{1}{c|}{61.19} & 122.37 & \multicolumn{1}{c|}{18.87} & \multicolumn{1}{c|}{37.75} & 75.49 \\ \hline
TVLSI'23\cite{Tan-TVLSI'23} & FP-32/16 & 59.3 & 0.013 & 2.22 & \multicolumn{1}{c|}{74.87} & \multicolumn{1}{c|}{149.75} & N/A & \multicolumn{1}{c|}{341.54} & \multicolumn{1}{c|}{683} & N/A \\ \hline
TVLSI'22\cite{FP-MPE} & FP-64/32/16 & 29.3 & 0.013 & 1.43 & \multicolumn{1}{c|}{97.61} & \multicolumn{1}{c|}{390} & N/A & \multicolumn{1}{c|}{220} & \multicolumn{1}{c|}{880} & N/A \\ \hline
TCAS-II'24\cite{Li-TCASII'24} & FP-32/16/INT8 & 15.86 & 0.011 & 1.47 & \multicolumn{1}{c|}{185.37} & \multicolumn{1}{c|}{741.49} & 1668.35 & \multicolumn{1}{c|}{294} & \multicolumn{1}{c|}{1176} & 2646 \\ \hline
Proposed Q-MAC & AdFxP-32/16/8 & 7.69 & 0.008 & 1.53 & \multicolumn{1}{c|}{397.92} & \multicolumn{1}{c|}{1591.68} & 6366.71 & \multicolumn{1}{c|}{392.31} & \multicolumn{1}{c|}{1569.23} & 6276.92 \\ \hline
\end{tabular}}
\vspace{-4mm}
\end{table*}

\subsection{Optimized Compute Unit : Q-MAC}
Efficient execution of matrix-vector multiplications and element-wise arithmetic is central to RL workloads, necessitating a runtime-configurable, precision-aware SIMD Q-MAC unit. The proposed Q-MAC achieves a favorable resource–throughput trade-off, leveraging a low-power compute (LPC) design philosophy~\cite{Li-TCASII'24, Tan-TVLSI'23} in contrast to traditional hybrid precision scaling (HPS)~\cite{Tan-TCAD'24, Tan-TCASII'24} and bit-partitioning techniques~\cite{Li-TCASII'22}. Q-MAC supports 8-, 16-, and 32-bit SIMD modes using a scalable array of 16 parallel 8-bit multipliers, with inspiration by adaptive fixed-point (AdFxP) methods\cite{flex-PE}. As shown in Fig.~\ref{fig:Q-MAC}, the architecture consists of an input handler that maps operands dynamically based on bit-width, followed by a shift–add adder tree with dedicated alignment and accumulation stages. A configurable output handler finalizes the results. It was observed that replacing the standard multiplier in the Q-MAC unit with approximate alternatives : such as Iterative Log Multiplier (ILM)~\cite{ILM_Ratko-TCASI'22}, CORDIC-based multiplier~\cite{GR_ACM_TRETS}, Posit-mult~\cite{lpre}, and Quant-MAC~\cite{QuantMAC}can maintain a quality of results (QoR) between 98.4\% and 99.2\%, while achieving significant resource savings. These implementations are highly scalable, leveraging the Q-MAC's architecture of 16 parallel 8-bit multiplier blocks. Specifically, CORDIC-mult and Quant-MAC achieve up to 42\% resource reduction with a modest 1.8\% accuracy loss. ILM provides 23\% savings at the cost of 1.2\% degradation, whereas Posit-mult offers a balanced design point with only 0.2\% accuracy drop, validating the Q-MAC’s flexibility and area efficiency with approximate quantized RL workloads.

Table \ref{fpga-mac} compared the FPGA utilization for proposed Q-MAC against existing MAC architectures. The proposed adaptive fixed-point based Q-MAC achieves significantly higher frequency without contributing much to power consumption compared to SoTA SIMD MAC units. It can be deduced that the significant LUTs, FFs resource savings is possible, with key-driver being fixed-point quantized implementation, compared to resource-heavy Floating point computations. This makes the design better suitable for FPGA-based resource-constrained Edge-AI applications. For ASIC-translation impact, we have discussed ASIC performance metrics in Table \ref{MAC-ASIC} using TSMC 28nm CMOS technology. The proposed Q-MAC provides significantly increased energy efficiency (up to 6.37 TOPS/W at 8-bit) and high compute density (up to 6.28 TOPS/mm² at 8-bit), outperforming prior SoTA designs in terms of both power and area efficiency.

\begin{figure}[!t]
    \centering
    \includegraphics[width=+++++\linewidth]{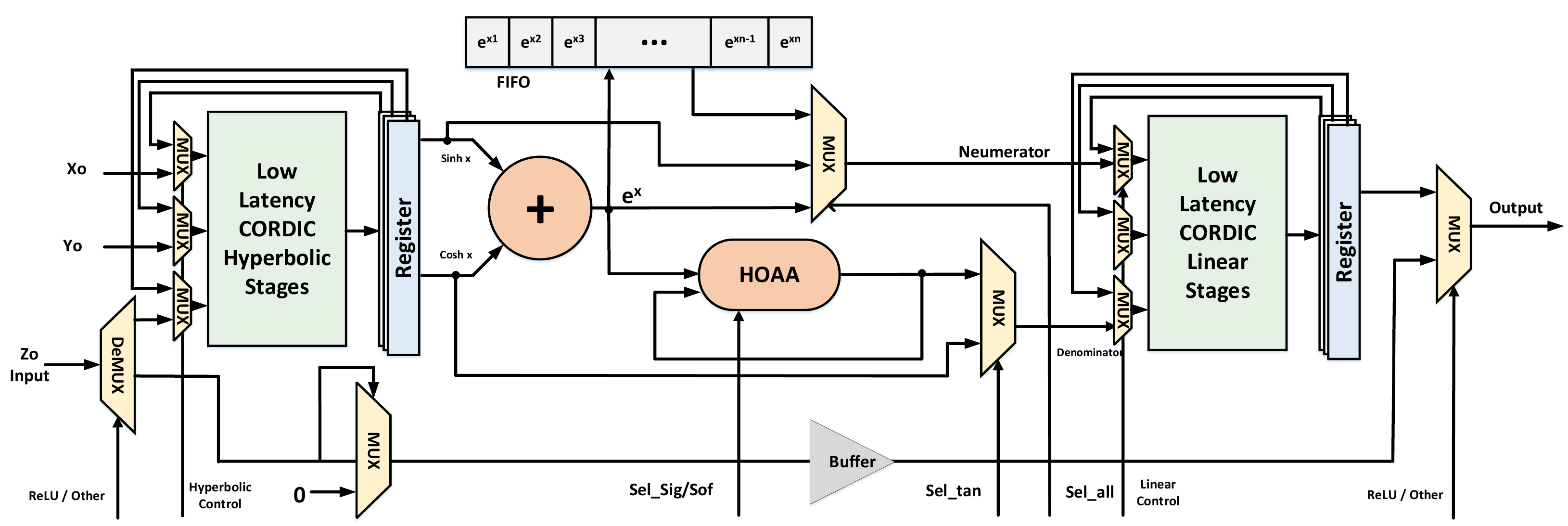}
    \caption{low-latency V-ACT architecture.}
    \label{fig:V-ACT}
    \vspace{-5mm}
\end{figure}

\begin{table}[!b]
\vspace{-5mm}
\caption{Resource utilization: V-ACT \cite{flex-PE, RNN-AF-TAI25, Trans-AF-TVLSI23}}
\label{AF-resources}
\renewcommand{\arraystretch}{1.2}
\resizebox{\columnwidth}{!}{%
\begin{tabular}{|ccccccc|}
\hline
\multicolumn{1}{|c|}{} & \multicolumn{3}{c|}{\textbf{TVLSI'25\cite{flex-PE}}} & \multicolumn{1}{c|}{\textbf{TVLSI'23\cite{Trans-AF-TVLSI23}}} & \multicolumn{1}{c|}{\textbf{TC'23\cite{LSTM-AF-TC'23}}} & \textbf{V-ACT} \\ \hline
\multicolumn{1}{|c|}{\textbf{NAF}} & \multicolumn{1}{c|}{Softmax} & \multicolumn{1}{c|}{Sigmoid} & \multicolumn{1}{c|}{Tanh} & \multicolumn{1}{c|}{Softmax} & \multicolumn{1}{c|}{Sigmoid/Tanh} & \begin{tabular}[c]{@{}c@{}}Tanh/Sigmoid/\\ Softmax/ReLU\end{tabular} \\ \hline
\multicolumn{1}{|c|}{Precision} & \multicolumn{3}{c|}{FP32} & \multicolumn{1}{c|}{16-b} & \multicolumn{1}{c|}{16-b} & SIMD 8/16/32 \\ \hline
\multicolumn{7}{|c|}{\textbf{FPGA utilization}} \\ \hline
\multicolumn{1}{|c|}{\textbf{FPGA}} & \multicolumn{3}{c|}{VC707} & \multicolumn{1}{c|}{Zynq-7} & \multicolumn{1}{c|}{Pynq Z1} & VC707 \\ \hline
\multicolumn{1}{|c|}{LUTs} & \multicolumn{1}{c|}{3217} & \multicolumn{1}{c|}{5101} & \multicolumn{1}{c|}{4298} & \multicolumn{1}{c|}{1215} & \multicolumn{1}{c|}{2395} & 763 \\ \hline
\multicolumn{1}{|c|}{FFs} & \multicolumn{1}{c|}{NR} & \multicolumn{1}{c|}{NR} & \multicolumn{1}{c|}{NR} & \multicolumn{1}{c|}{1012} & \multicolumn{1}{c|}{1503} & 1058 \\ \hline
\multicolumn{1}{|c|}{Delay (ns)} & \multicolumn{1}{c|}{91.94} & \multicolumn{1}{c|}{109} & \multicolumn{1}{c|}{56.6} & \multicolumn{1}{c|}{332} & \multicolumn{1}{c|}{21} & 9.28 \\ \hline
\multicolumn{1}{|c|}{Power (mW)} & \multicolumn{1}{c|}{115} & \multicolumn{1}{c|}{121} & \multicolumn{1}{c|}{135} & \multicolumn{1}{c|}{165} & \multicolumn{1}{c|}{125} & 48.56 \\ \hline
\multicolumn{1}{|c|}{\begin{tabular}[c]{@{}c@{}}Arith.   Intensity \\      (nJ/op)\end{tabular}} & \multicolumn{1}{c|}{10.57} & \multicolumn{1}{c|}{13.2} & \multicolumn{1}{c|}{7.63} & \multicolumn{1}{c|}{547} & \multicolumn{1}{c|}{2625} & 0.7 \\ \hline
\multicolumn{7}{|c|}{\textbf{ASIC Utilization}} \\ \hline
\multicolumn{1}{|c|}{Tech. (nm)} & \multicolumn{3}{c|}{28} & \multicolumn{1}{c|}{28} & \multicolumn{1}{c|}{45} & 28 \\ \hline
\multicolumn{1}{|c|}{Area (um2)} & \multicolumn{1}{c|}{41536} & \multicolumn{1}{c|}{2234} & \multicolumn{1}{c|}{5060} & \multicolumn{1}{c|}{3819} & \multicolumn{1}{c|}{24608} & 8236 \\ \hline
\multicolumn{1}{|c|}{Delay (ns)} & \multicolumn{1}{c|}{5.95} & \multicolumn{1}{c|}{7.58} & \multicolumn{1}{c|}{3.92} & \multicolumn{1}{c|}{1.6} & \multicolumn{1}{c|}{NR} & 2.1 \\ \hline
\multicolumn{1}{|c|}{Power (mW)} & \multicolumn{1}{c|}{74.8} & \multicolumn{1}{c|}{10.06} & \multicolumn{1}{c|}{8.75} & \multicolumn{1}{c|}{1.58} & \multicolumn{1}{c|}{151} & 4.67 \\ \hline
\multicolumn{1}{|c|}{PDP (pJ)} & \multicolumn{1}{c|}{445.06} & \multicolumn{1}{c|}{76.2548} & \multicolumn{1}{c|}{34.3} & \multicolumn{1}{c|}{2.528} & \multicolumn{1}{c|}{NR} & 9.8 \\ \hline
\end{tabular}}
\end{table}

\begin{table*}[!t]
\caption{Comparative analysis for FPGA utilization with SoTA AI Architectures}
\label{FPGA-util-comp-arch}
\renewcommand{\arraystretch}{1.15}
\resizebox{\textwidth}{!}{%
\begin{tabular}{|c|cc|c|c|c|c|c|c|c|c|c|}
\hline
\textbf{} & \multicolumn{2}{c|}{\textbf{E2HRL\cite{E2HRL}}} & \textbf{Synergy\cite{Synergy}} & \textbf{TVLSI'22\cite{TVLSI22-SSD}} & \textbf{TCAS-I'22\cite{DTN-TCASI'22}} & \textbf{TCAD'23\cite{WJ-TCAD'23}} & \textbf{TVLSI'23\cite{WL-TVLSI'23}} & \textbf{TCAS-I'24\cite{Wu-TCASI'24}} &  \multicolumn{2}{c|}{\textbf{Proposed}} \\ \hline
\textbf{Application} & \multicolumn{1}{c|}{\textbf{FC-HRL}} & \textbf{LSTM-HRL}  & \textbf{DNN} & \textbf{\begin{tabular}[c]{@{}c@{}}RealTime\\      SSDLite\end{tabular}} & \textbf{YOLO v3} & \textbf{MobileNet-V2} & \textbf{Yolo} & \textbf{YOLO} & \textbf{Q-FC} & \textbf{Q-LSTM} \\ \hline
\textbf{FPGA} & \multicolumn{1}{c|}{Artix7} & Artix7 & Zynq & Arria 10 & KCU15 & ZCU102 & Zynq US+ & Nexys A7-100T & Virtex-7 & Virtex-7 \\ \hline
\textbf{I/P size} & \multicolumn{1}{c|}{40 × 30 × 3} & 40 × 30 × 3 & 32 × 32 × 3 & NR & 224x224 & 224x224x3 & 416x416x3 & 320×320×3 & 32 x 32 x 3 & 32 x 32 x 3 \\ \hline
\textbf{Bit-width} & \multicolumn{1}{c|}{32b} & 32b & 32b & FxP-16b & 16-b & FxP-8b & Q-8b & Q-8b & Q-8/16/32 & Q-8/16/32 \\ \hline
\textbf{Freq (MHz)} & \multicolumn{1}{c|}{100} & 100  & 100 & 200 & 200 & 333 & 300 & 100 & 250 & 250 \\ \hline
\textbf{Proc. (FPS)} & \multicolumn{1}{c|}{\textbf{1110}} & 435  & 63.5 & 84.8 & 25.5 & \textbf{1910} & 64.5 & 76.75 & \textbf{2835} & 924 \\ \hline
\textbf{Throughput} & \multicolumn{1}{c|}{4} & 1  & 1.67 & 1.45 & 0.675 & 1.225 & \textbf{9.95} & 0.396 & \textbf{11.2} & 2.8 \\ \hline
\textbf{Power (mW)} & \multicolumn{1}{c|}{\textbf{348}} & 389 & 1015 & 980 & 21090 & NR & 6580 & 2203 & \textbf{196} & \textbf{256} \\ \hline
\textbf{Energy (mJ/frame)} & \multicolumn{1}{c|}{\textbf{0.3}} & 0.8  & 33.7 & 0.35 & NA & NR & 9.8 & NR & 0.15 & \textbf{0.26} \\ \hline
\textbf{\begin{tabular}[c]{@{}c@{}}Energy Efficiency \\      (GOPS/W)\end{tabular}} & \multicolumn{1}{c|}{\textbf{11.4}} & 3 & 1.65 & 1.48 & 3.21 & NR & 1.5 & 43 & \textbf{26.1} & \textbf{7.8} \\ \hline
\end{tabular}}
\vspace{-4mm}
\end{table*}

\subsection{Unified AF: V-ACT}

To support diverse activation function (AF) requirements across different layers : such as ReLU in Q-Conv, Softmax in classification, and Tanh/Sigmoid in Q-LSTM - we propose a Versatile Activation Function Unit (V-ACT). Prior works like Flex-SFU~\cite{Flex-SFU} and Flex-PE~\cite{flex-PE} have explored Taylor-series and CORDIC-based AFs respectively. However, Taylor-based methods suffer from interpolation granularity issues and convergence sensitivity, making them less resource-efficient. While Flex-PE supports MAC and AF across multiple precisions (FxP4/8/16/32), it introduces performance degradation and is less effective for specialized RL workloads. We used similar philosophy with low-latency CORDIC-hyperbolic stages\cite{low-latency-cordic} to formulate Versatile ACT which support ReLU, Sigmoid, Tanh, and Softmax with FxP8/16/32 precision. The low-latency CORDIC approach is resource-efficient and converges in (3n/8  + 1) cycles compared to (n/2 +1 ) stages in unified CORDIC approach\cite{GR_ACM_TRETS}. The detailed V-ACT architecture is illustrated in Fig. \ref{fig:V-ACT}, with control logic enabled precision selection and computation through configurable data paths. The HOAA-enabled approach enhanced the performance by 21\%. The use of FIFO for exponent buffering and functional separation between hyperbolic \& linear paths enhances pipeline throughput and scalability, making it highly suitable for parallel inference engines. The FPGA and ASIC resources for the proposed V-ACT architecture are reported in Table \ref{AF-resources} and compared against recent SoTA designs. The significant improvements are observed, credit attributed to resource-efficient CORDIC approach. We have observed significant resource savings up-to 15\% with similar Flex-PE\cite{flex-PE} and up-to 50\% compared to 32-bit Flex-SFU\cite{Flex-SFU} at TSMC 28nm. The key focus of our approach is on minimal delay effectively contributing to higher performance (TOPS) and low-arithmetic intensity leading to improved energy efficiency. The reconfigurable nature with versatile AF and precision itself emphasizes on reduced FPGA resource utilization or compute density.

\subsection{Architectural Integration \& Analysis}
The experimental setup for evaluating the proposed methodology includes a software-based Q-Actor analysis and hardware-oriented architectural emulation to verify the RL accelerator codesign. The FPGA synthesis and implementation was carried out with the advanced micro devices (AMD) Vivado Design Suite, and post-implementation resource utilization was reported. The architectural integration for Q-Force RL was done and verified with UART-based TeraTerm Receiver with Virtex VC707 FPGA board and the FPGA implementation results are reported in Table \ref{FPGA-util-comp-arch}, showing significant reduction in resource utilization (LUTs and FFs),power consumption, and delay when compared to SoTA works. The Q-MAC unit and V-ACT unit was synthesized using CMOS 28-nm technology with the Synopsys Design Compiler, and post-synthesis parameters are reported as mentioned in previous subsections, respectively. We have also re-implemented SoTA designs with the similar parameters for a fair comparison.

The SISD implementation for the E2HRL\cite{E2HRL} architecture with FxP32 was implemented on mobile CPUS such as dual-core Denver and quad-core Arm Cortex-A57 at frequency of 345, 2035 MHz. The latency necessary to achieve 30 FPS target is 33.33 mS. We re-implemented our design with Arm-Neon SIMD type Ryzen 7 6800H processor at 2035 MHz. We observed the performance improvement in CPU Latency (6.2 ms for 32-bit) and around 2.6 $\times$ for 8-bit implementation and 1.4$\times$ for 16-bit implementation compared to 32-bit implementation. The energy efficiency (1.42 GOP/s/W for 32-bit) was observed approximately and around 1.7 $\times$ and 3.1 $\times$ for 16-bit and 8-bit configuration for Q-FC and Q-LSTM HRL configurations, which was derived from CPU Power(mW). The similar gains can be observed in performance (GOP/s), and energy (mJ) was also noticed. 

An architectural comparison of various FPGA-based implementations is reported in Table \ref{FPGA-util-comp-arch}., where the proposed Q-FC and Q-LSTM demonstrate significant improvements with high throughput (2835 FPS for Q-FC) and low energy consumption (0.15 mJ/frame), resulting in excellent power efficiency of 26.1 GOPS/W for Q-FC and 7.8 GOPS/W for Q-LSTM substantially outperforming prior works in both speed and energy efficiency. The FPGA implementation can be further analysed with different number of PEs from 1 to 8 at 232 MHz. The best configuration, with 8 PEs, achieved a peak performance for Q-FC model achieved  11 GOPS at 196 mW and Q-LSTM reached 2.8 GOPS at 256 mW. Rge energy consumption saturated around 23 mJ/frame. The models also highlight the resource efficiency and scalability of the QForce-RL architecture at both FPGA and mobile platforms in diverse configurations.

\section{Conclusion \& Future work}
\label{sec4}
In this paper, we presented \textbf{QForce-RL}, an FPGA-optimized, quantized RL compute engine designed for efficient and scalable deployment. By leveraging precision-aware SIMD \textbf{Q-MAC} and a low-latency, reconfigurable CORDIC-based (\textbf{V-ACT}), the proposed architecture significantly reduces computational complexity, memory footprint, and communication overhead without compromising RL model performance. Empirical evaluations on FPGA and mobile CPU platforms demonstrate substantial improvements in throughput, energy efficiency, and latency for both Q-FC and Q-LSTM configurations, positioning QForce-RL as a robust and resource-efficient solution for edge deployments.

While this work primarily targets RL, the proposed compute blocks: Q-MAC and V-ACT are also applicable to general DNN workloads. Thus, by integration of these units to achieve functional equivalent of CORDIC-Neuron\cite{GR_ACM_TRETS} or Flex-PE\cite{flex-PE} based Systolic-array/Layer-multiplexed architecture provides significantly enhanced throughput outperforming SoTA DNN works. Incorporating Q-MAC and V-ACT into a DNN accelerator achieves an inference latency of 112~ms (69.3~GOPS), outperforming LPRE~\cite{lpre} (184~ms, 42.3~GOPS), CORDIC-Neuron~\cite{GR_ACM_TRETS} (772~ms, 4.95~GOPS), and Jetson Nano (226~ms, 34.38~GOPS), with nearly 1\% better application accuracy. The future scope includes integration into mixed-precision, layer-wise quantized AI accelerators.

\bibliographystyle{ieeetr}
\bibliography{conference}

\end{document}